\documentclass[sigconf]{acmart}
\AtBeginDocument{%
  }

\setcopyright{acmlicensed}
\copyrightyear{2018}
\acmYear{2018}
\acmDOI{XXXXXXX.XXXXXXX}
\acmConference[Conference acronym 'XX]{Make sure to enter the correct
  conference title from your rights confirmation email}{June 03--05,
  2018}{Woodstock, NY}
\acmISBN{978-1-4503-XXXX-X/2018/06}

\usepackage{pgfplots}
\usepackage{xcolor}
\usepackage{geometry}
\usepackage{balance}
\copyrightyear{2025}
\acmYear{2025}
\setcopyright{acmlicensed}\acmConference[MM '25]{Proceedings of the 33rd ACM International Conference on Multimedia}{October 27--31, 2025}{Dublin, Ireland}
\acmBooktitle{Proceedings of the 33rd ACM International Conference on Multimedia (MM '25), October 27--31, 2025, Dublin, Ireland}
\acmDOI{10.1145/3746027.3754584}
\acmISBN{979-8-4007-2035-2/2025/10}

\begin{document}

\title{VisAug: Facilitating Speech-Rich Web Video Navigation and Engagement with Auto-Generated Visual Augmentations}

\author{Baoquan Zhao}
\affiliation{%
  \institution{Sun Yat-sen University}
  \city{Zhuhai}
  \country{China}
  }  

  \author{Xiaofan Ma}
\affiliation{%
  \institution{Sun Yat-sen University}
  \city{Guangzhou}
  \country{China}}  
    \author{Qianshi Pang}
\affiliation{%
  \institution{Sun Yat-sen University}
  \city{Zhuhai}
  \country{China}}  
  \author{Ruomei Wang}
\affiliation{%
  \institution{Sun Yat-sen University}
  \city{Zhuhai}
  \country{China}}

\author{Fan Zhou}
\affiliation{%
  \institution{Sun Yat-sen University}
  \city{Guangzhou}
  \country{China}}  
  
\author{Shujin Lin}
\authornote{Corresponding author. E-mail: linshjin@mail.sysu.edu.cn}
\affiliation{%
  \institution{Sun Yat-sen University}
  \city{Guangzhou}
  \country{China}}

\renewcommand{\shortauthors}{Baoquan Zhao et al.}

\begin{abstract}
The widespread adoption of digital technology has ushered in a new era of digital transformation across all aspects of our lives. Online learning, social, and work activities, such as distance education, videoconferencing, interviews, and talks, have led to a dramatic increase in speech-rich video content. In contrast to other video types, such as surveillance footage, which typically contain abundant visual cues, speech-rich videos convey most of their meaningful information through the audio channel. This poses challenges for improving content consumption using existing visual-based video summarization, navigation, and exploration systems. In this paper, we present VisAug, a novel interactive system designed to enhance speech-rich video navigation and engagement by automatically generating informative and expressive visual augmentations based on the speech content of videos. Our findings suggest that this system has the potential to significantly enhance the consumption and engagement of information in an increasingly video-driven digital landscape.
\end{abstract}



\begin{teaserfigure}
  \includegraphics[width=\textwidth]{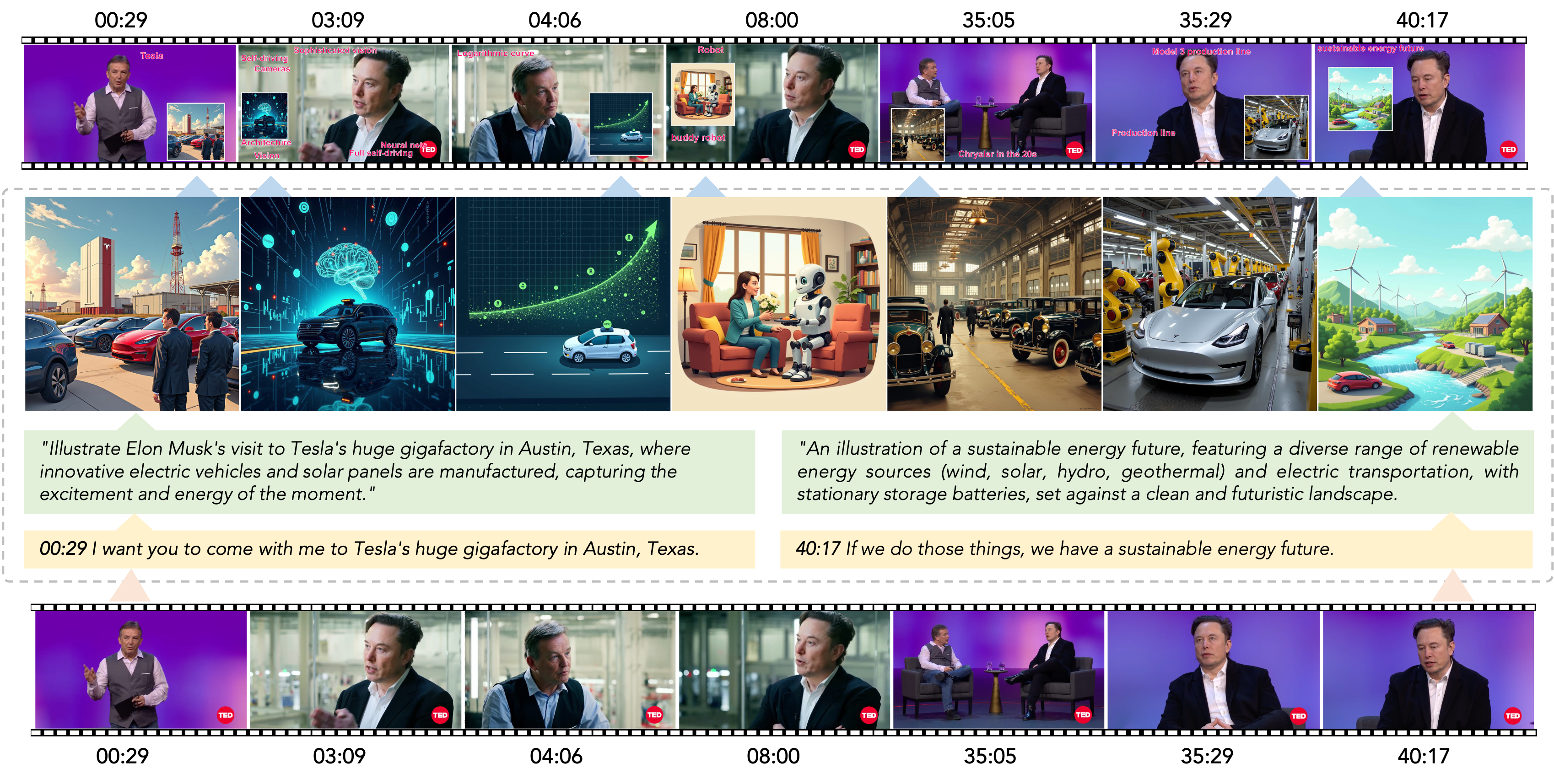}
  \caption{Overview of speech-rich web video augmentation method. Bottom row shows original TED video frames; top row shows frames with auto-generated visual augmentations. The middle section demonstrates the augmentation process at timestamps 00:29 and 40:17, where the system evaluates transcript imageability, generates context-aware text-to-image prompts, and integrates resulting image augmentations with key phrases into less prominent frame areas to facilitate content navigation.
}
  \Description{}
  \label{fig:teaser}
\end{teaserfigure}

\maketitle
\section{Introduction}
The vast transformation led by the continuously evolving digital age has changed the way verbal communication is conducted from monomodality to more effective and engaging multimodality, multisensory representations~\cite{metatla2016tap,dodani2022birdbox,rubab2023exploring}.
As a consequence, we have witnessed the rapid proliferation of speech-rich video recordings over the Internet derived from a broad spectrum of activities such as online meetings, lectures, talks, interviews, discussions, conversations, narrations, storytelling, and beyond~\cite{zhao2016new,zhao2019new}. As opposed to visually-informative counterparts like surveillance footage, whose frames themselves convey abundant visual clues with which people can identify content of interest with ease, the consumption of such kind of video content remains an ongoing challenge due to the lack of sufficient visual aids except for talking head shots. Despite the increasing interest in developing tools to facilitate video content navigation and digestion, the inherent difference between visually-informative and speech-rich videos makes it difficult to providing internet users with instant gratification using existing solutions that have scarcely been optimised for speech-rich videos.

\begin{figure}[!t]
    \centering
    \includegraphics[width=0.48\textwidth]{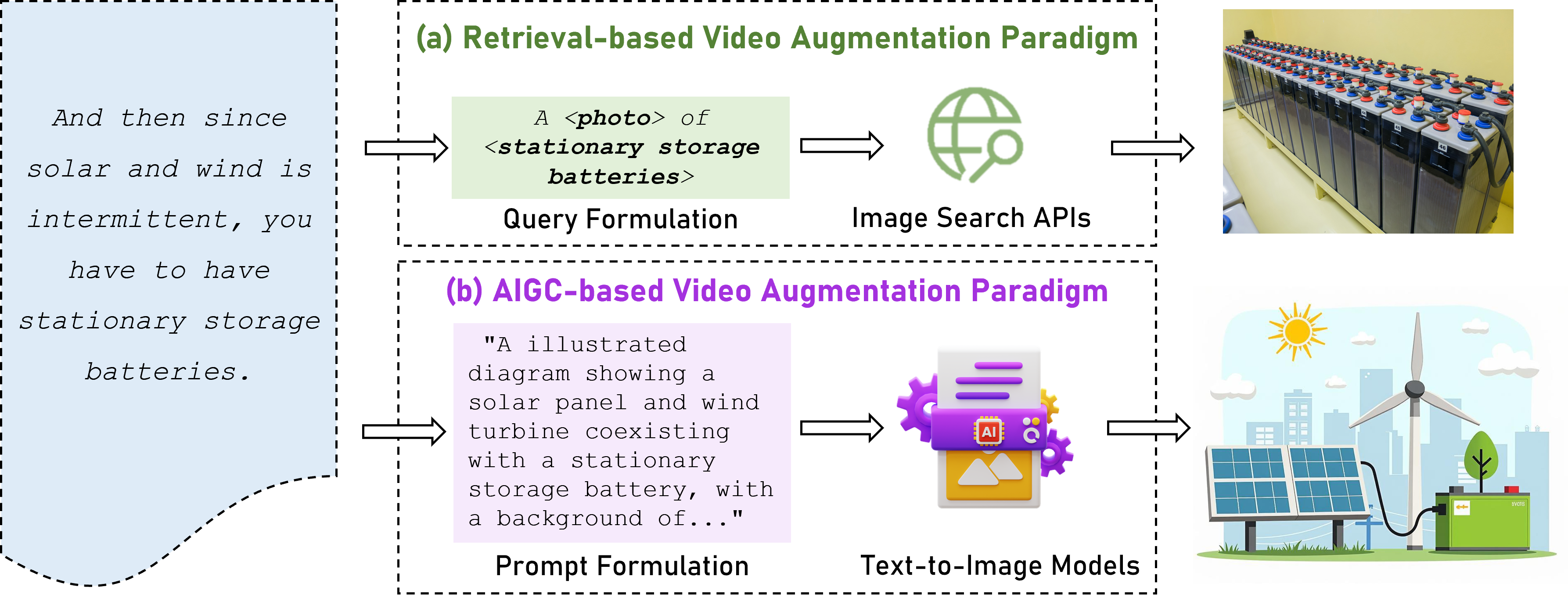}
 \caption{Comparison of video augmentation paradigms: (a) retrieval-based methods use image search APIs to find relevant images from online repositories based on natural language queries; (b) our AIGC-based approach leverages text-to-image models to generate contextual visual aids directly from processed text prompts.}
    \label{fig:paradigm}
\end{figure}

To alleviate the situation, previous studies have probed the video augmentation pathway of suggesting relevant images as visual aids from either local photo repositories or online image databases. For instance, a recent work~\cite{liu2023visual} introduced a real-time system for video conferencing content augmentation by extracting textual queries from video captions and then retrieving relevant images from a local photo album or online repository using the Bing image search API. Other studies in this vein have also explored using image retrieval to augment other forms of media, such as travel podcasts~\cite{xia2020crosscast}. While recent progress paves the way for the development of retrieval-based content augmentation, existing solutions still fall far short of content relevance and expressiveness of visual aids, especially for complex and nuanced queries. As illustrated in Figure~\ref{fig:paradigm}, given a piece of text, the suggested visual content using existing retrieval-based methods is far from meeting expectations with regard to multifaceted representation. 
In addition, complex, abstract, or novel ideas that may not have direct visual counterparts in existing databases cannot be effectively visualized through retrieval-based methods.
In contrast, this study seeks to take advantage of the well-developed artificial intelligence-generated content (AIGC) infrastructure to explore the potential of automatically generating images as visual aids from natural language extracted from a speech-rich video to fuel interest in AIGC-based video augmentation, which can serve as a new enabler for more effective verbal communication and engagement beyond the reach of the retrieval-based paradigm. For instance, a single AI-generated image can incorporate multiple elements or concepts discussed in the transcript, creating a rich, composite visual that encapsulates the full breadth of the content.

Towards this end, the first task is to automatically measure the imageability (a.k.a. visualness) of text, a concept from psycholinguistics that examines the link between language and mental visualization~\cite{verma2023learning}, i.e., how well a piece of text, whether a single word or a sentence, triggers a mental image in the reader. For instance, the words like \textit{sunset, mountain} or sentences like \textit{the sun-drenched beach stretched out before them, the turquoise water lapping gently against the pearly white sand.} are easy to evoke a vivid image in people's minds, making them highly imageable. while words like \textit{loyalty, freedom} or sentences like \textit{because of her honesty, she was trusted by everyone.} are more abstract and don't lend themselves as easily to visualization, resulting in lower imageability. While enormous efforts have gone into developing human-assigned word-level imageability databases and computational methodologies to automatically estimate a word's visualness based on its properties, existing solutions either fall short of limited word coverage or aggregation issues, i.e., combining the imageability scores of individual words directly doesn't necessarily reflect the visualness of a sentence. In addition, as the video content evolves, it is necessary to generate images that reflect subtle changes or progressions in topics, something static retrieval-based methods might struggle with.
Understanding how entire sentences or phrases evoke mental imagery, especially under the circumstances of the utterance of speech addressed in this work, is highly desired but still in its infancy.

Crafting a clear and concise textual prompt is critical for today's text-to-image generation models to create stunning and creative images. Once a piece of text is identified as imageable, the second task is to tailor a piece of text as a proper prompt for generative AI models to achieve desired visual outcomes. Intuitively, one may raise a question about whether a text-to-image model can take the whole original text without modification or refinement as a prompt for image generation. Unfortunately, such a naive scheme tends to fail to meet expectations since it may come with a lack of focus issue, that is, a full sentence might contain unnecessary details that distract the model from the core concept to be visualized. Furthermore, compared to written language, spoken language in speech-rich videos may also have the issues of ambiguities: Words can have multiple meanings depending on context, and sentences may lack clear structure.
and this may lead to misinterpreted intentions or irrelevant outcomes.  Robust and effective pathways for automated prompt creation from text remain largely under-explored. 

In this paper, we introduce VisAug, a new web-based intelligent system designed to enhance speech-rich videos through auto-generated visual augmentations. Our approach leverages a dual-channel processing pipeline that simultaneously analyzes visual and textual (derived from audio) content. The visual channel employs saliency detection and segmentation to identify key areas in video frames, while the textual channel utilizes speech recognition and advanced natural language processing techniques, including text imageability assessment and keyphrase extraction. These processes culminate in the generation of context-aware image augmentations and the strategic placement of both images and keyphrases in non-salient regions of the video frames. By seamlessly integrating these augmentations, VisAug aims to improve content navigation, comprehension, and viewer engagement without obscuring original video content.  It is worth noting that the proposed method and the resultant visual aids can also be used to augment written languages in various forms such as media posts, reports, scripts, novels, poems, essays, and beyond.
The major contributions of this work are summarized as follows:
\begin{itemize}
    \item We introduce a novel AIGC-powered interactive visual augmentation system for facilitating speech-rich video content navigation, exploration, and engagement. To the best of our knowledge, this is a pioneering attempt towards augmenting speech-rich videos with AI-generated images as visual aids.
    \item We propose a novel context-aware prompt formulation scheme that leverages text-to-image generation models to create faithful and expressive visual augmentations. It considers both global video context and local segment relevance, ensuring that generated images are not only visually compelling but also semantically aligned with the video content. 
    \item Experiment results demonstrate the superiority and versatility of the proposed system over existing web video user interfaces.
\end{itemize}

\begin{figure*}[!t]
    \centering
    \includegraphics[width=\textwidth]{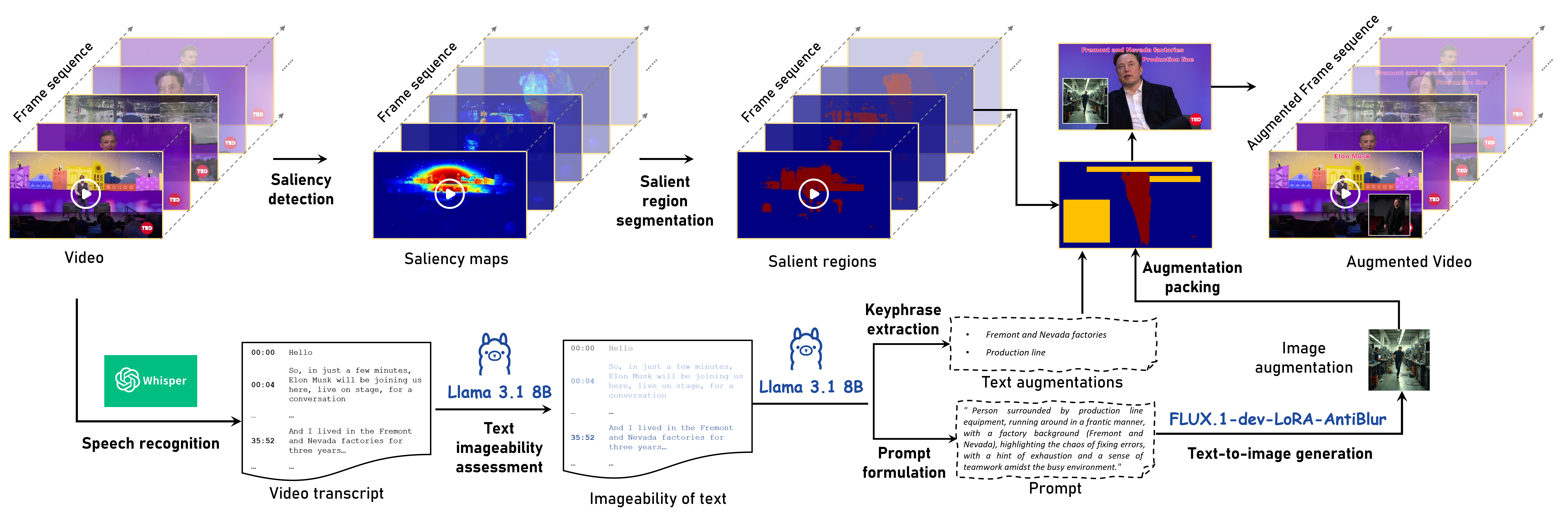}
    \caption{Pipeline of the proposed method.}
    \label{fig:pipeline}
\end{figure*}

\section{Related Work}
\subsection{Text-to-Visual Augmentation Systems}
Text-to-visual augmentation enhances content navigation across text~\cite{hoffswell2018augmenting,goffin2020interaction,masson2023charagraph,leake2020generating,ahn2023story}, audio~\cite{leake2020generating,xia2020crosscast}, and video~\cite{chen2022sporthesia,liu2023visual,shen2023data,zhao2017novel,xu2022semantic}. Previous work~\cite{leake2020generating} explored generating audio-visual slideshows from text by extracting keywords to retrieve relevant images. Crosscast~\cite{xia2020crosscast} enhanced travel podcast consumption by synchronizing visual aids with audio narration through geographic location identification.
Visual Caption~\cite{liu2023visual} enriches verbal communication with on-the-fly visuals in open-vocabulary conversations. However, it faces limitations with rare queries and retrieval-based visuals that may lack context. Our system addresses these challenges by leveraging AIGC infrastructure to create expressive visuals, representing a pioneering approach to verbal communication augmentation.

\subsection{Automatic Text Imageability Assessment}
Text imageability influences communication and cognition. Wolfe et al.~\cite{wolfe2019theoretically} found that higher imageability leads to verbatim representations, while Ostryn et al.~\cite{ostryn2021text} explored imageability effects in alternative communication for autism spectrum disorders. 
Traditional imageability ratings relied on human evaluation~\cite{wilson1988mrc,liu2014automatic}, but computational methods have emerged~\cite{kao2015computational,ljubevsic2018predicting,kastner2020estimating}, primarily focused on word-level assessment~\cite{matsuhira2020imageability}. Recent research has leveraged text-to-image models~\cite{wu2023composition} and vision-language models~\cite{verma2023learning} for sentence-level imageability. Our approach utilizes LLMs for context-aware sentence-level narrative visualness assessment, specifically designed for speech-rich video augmentation.

\subsection{Prompt Crafting from Natural Language}
AIGC advancements~\cite{hong2020f2gan,chen2023controlstyle,ding2022cogview2,qu2023layoutllm,wang2023semantic,chen2022re,li2023gligen} have enabled visual content creation through text-to-image models. Effective prompt crafting~\cite{zhang2021differentiable,ge2023expressive} is essential but often requires domain expertise and time-consuming tuning~\cite{zhou2022learning}.
Recent approaches include template selection without labeled examples~\cite{sorensen2022information} and automatic prompt engineering~\cite{zhou2022learning}. Despite progress, automatic prompt formulation from spoken language remains nascent. We systematically investigate prompt crafting patterns for visual aids generation and introduce a novel automated method to enhance speech-rich video augmentation through text-to-image generation.

\section{Proposed Method}
\subsection{Methodology Overview}
As shown in Figure~\ref{fig:pipeline}, the methodology of the proposed VisAug system is built on a dual-channel processing pipeline, leveraging both the visual and textual (derived from audio) content of web videos. The primary functions of the visual content processing branch are video frame saliency detection and salient region segmentation. These processes identify the most visually important or attention-grabbing areas within each video frame. This saliency information is crucial for the subsequent stage of visual augmentation packing, where it guides the placement and integration of generated visual elements. Running parallel to the visual analysis, the textual content processing branch begins with speech recognition to transcribe the audio content into text. The resulting transcript undergoes comprehensive natural language processing, including text imageability assessment, keyphrase extraction, and context-aware prompt formulation. The resultant prompt will be fed into a text-to-image model for image augmentation generation. In the visual augmentation packing stage, the system leverages the saliency maps from the visual branch to determine optimal placement and styling of the generated augmentations, including the extracted keyphrases and generated image. We detail each step of the proposed method in the following subsections.

\subsection{Visual Content Processing}
This module is meticulously designed to detect salient regions within video frames, playing a crucial role in informing the placement of generated visual augmentations. By identifying the most visually important or attention-grabbing areas of each frame, the system can determine optimal locations for placing additional visual elements without obscuring critical parts of the original content, thereby enhancing the content browsing experience. In our implementation, the MBD-based saliency detection method~\cite{zhang2015minimum} was employed in view of its effectiveness and efficiency. Recognizing that a transcript may span multiple shots or scenes, the system employs a bitwise saliency map cumulative technique to ensure the quality of subsequent visual augmentation placement. Specifically, let $t_s$ and $t_e$ be the start and end times of a transcript segment, respectively. We sample video frames at a rate of 1 frame per second, i.e., 
$\{F_i \mid i \in \mathbb{Z}, t_s \leq i \leq t_e\}$. For each sampled frame $F_i$, we detect its saliency map and then apply a binarization step to the saliency map using an adaptive thresholding scheme. The resultant binary saliency map is denoted as $S(F_i)$. The cumulative saliency map is then calculated through bitwise addition: 
\begin{equation}
    S_{\text{cumulative}} = \bigvee_{i=t_s}^{t_e} S(F_i),
\end{equation} where $\bigvee$ denotes the bitwise OR operation. The resulting $S_{\text{cumulative}}$ represents the final salient region map for the entire transcript segment, accounting for all shot transitions within the period $[t_s, t_e]$. This map can then be used to guide the placement of visual augmentations in non-salient areas across the duration of the transcript.

\subsection{Textual Content Processing}

\textbf{Speech recognition.} Given a speech-rich video, 
the system first attempts to harvest existing subtitles from the web. This prioritizes the use of professionally created or crowdsourced subtitles. If no subtitles are available, the system employs WhisperX~\cite{bain2023whisperx}, an advanced speech recognition tool, to convert the speech in the audio track into a time-aligned transcript. The resulting transcript, whether harvested or generated, serves as the foundation for subsequent processing steps, including text imageability assessment, keyphrase extraction, and text-to-image prompt formulation.

\begin{figure}[!t]
    \centering
\includegraphics[width=0.45\textwidth]{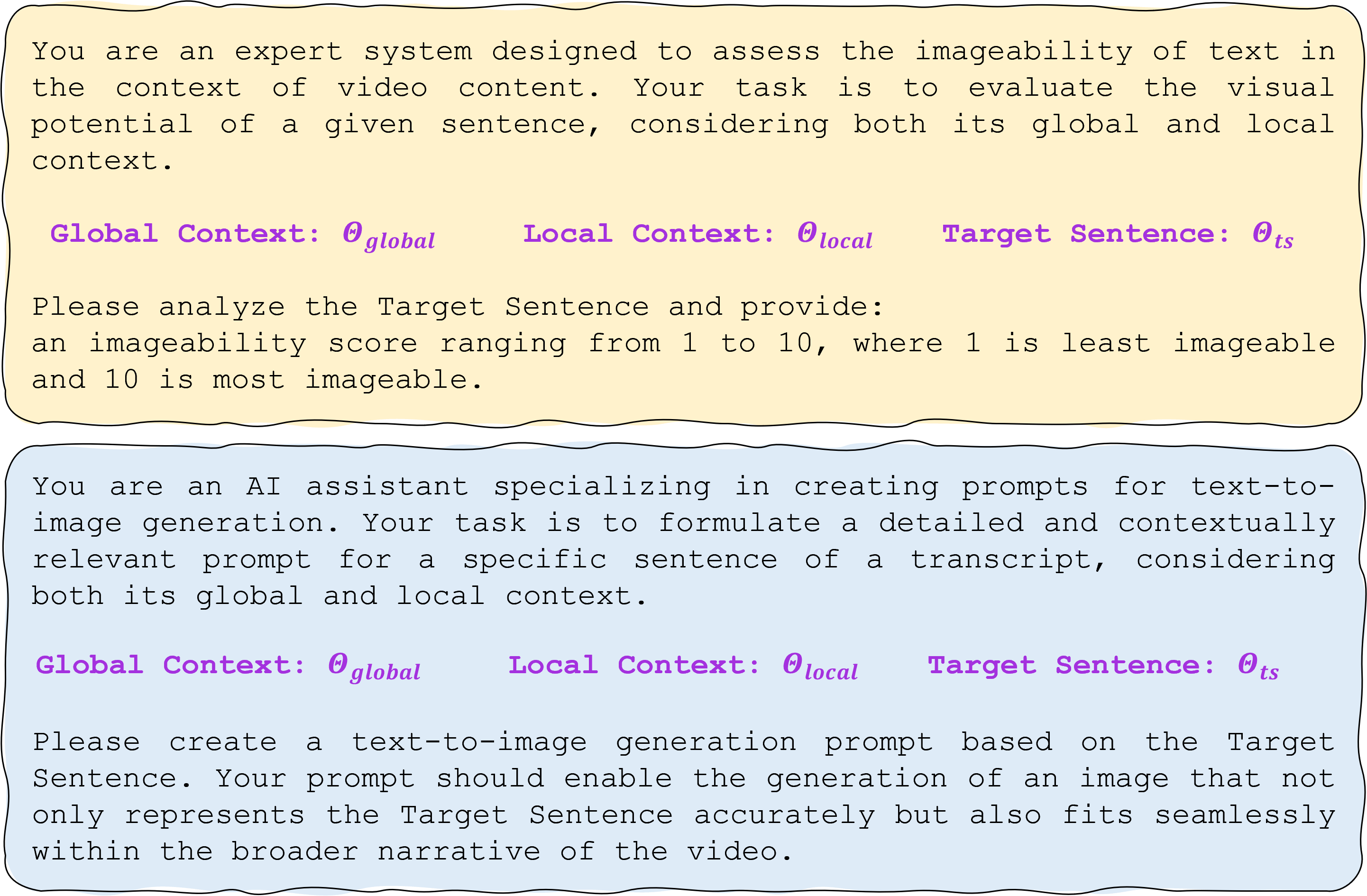}
    \caption{Prompts used for text imageability assessment (top) and text-to-image prompt formulation (bottom).}
    \label{fig:prompt}
\end{figure}

\textbf{Context-aware sentence-level text imageability assessment.} 
To address the limitations of traditional word-level imageability assessment methods, we leverage large language models to provide a more nuanced and comprehensive evaluation of the visual potential of spoken languages in speech-rich videos. This allows for a more holistic assessment that captures complex ideas, relationships, and scenarios that may not be apparent at the word level. In our implementation, the open-source Meta Llama 3.1 8B Instruct model~\cite{dubey2024llama} is used to perform this task. The 8B parameter version of Llama 3.1 represents a balance between computational efficiency and performance. It's large enough to handle complex language understanding tasks while being more manageable in terms of computational resources compared to larger models. In addition to the task description and the target sentence $\Theta_{ts}$, we integrate the global context  $\Theta_{global}$ and local context $\Theta_{local}$ into the prompt for text imageability assessment.  $\Theta_{global}$ enables the model to understand overarching themes, narrative structure, and key concepts that span the entire video. Using the entire video transcript as global context could indeed be overwhelming and potentially inefficient.  Instead, we carry out text summarization to distill the key information while maintaining the essence of the global context. For each target transcript segment, a set of adjacent transcripts is considered as local context. This allows the model to understand immediate contextual cues, transitions, and short-term thematic shifts that might influence the imageability of the target segment. In our implementation, we select the 5 segments before the target segment $\Theta_{ts}$ as the local context $\Theta_{local}$.  For each transcript segment, we fed the corresponding prompt (see Figure~\ref{fig:prompt}, top) in the Llama 3.1 8B Instruct model to harvest an imageability score from 1 to 10, where 1 is least imageable and 10 is most imageable.

\textbf{Context-aware prompt formulation for text-to-image generation.}
We filter all the transcript segments and select only those with an imageability score above 5 for further image augmentation. This threshold ensures that visual augmentations are generated only for content with high visual potential.
Similarly, we leverage the same contextual information used in text imageability assessment to formulate the text-to-image prompts 
 (see Figure~\ref{fig:prompt}, bottom). 
This context-aware approach ensures that the resulting prompts not only accurately represent the specific high-imageability segments but also maintain coherence with the video's broader narrative and immediate context. The Llama 3.1 8B Instruct model is then tasked with generating a text-to-image prompt. Finally, we employ a specialized text-to-image generation model, FLUX.1-dev-LoRA-AntiBlur, a functional LoRA trained on FLUX.1-dev, for creating image augmentations. 

\textbf{Text augmentation extraction.} 
In addition to the image augmentations, the module also conducts keyphrase extraction using the Llama 3.1 8B Instruct model, identifying significant concepts and topics within the transcript. These extracted keyphrases serve as text augmentations, highlighting important information and providing anchors for visual representation.

\begin{figure*}[!t]
    \centering
    \includegraphics[width=\textwidth]{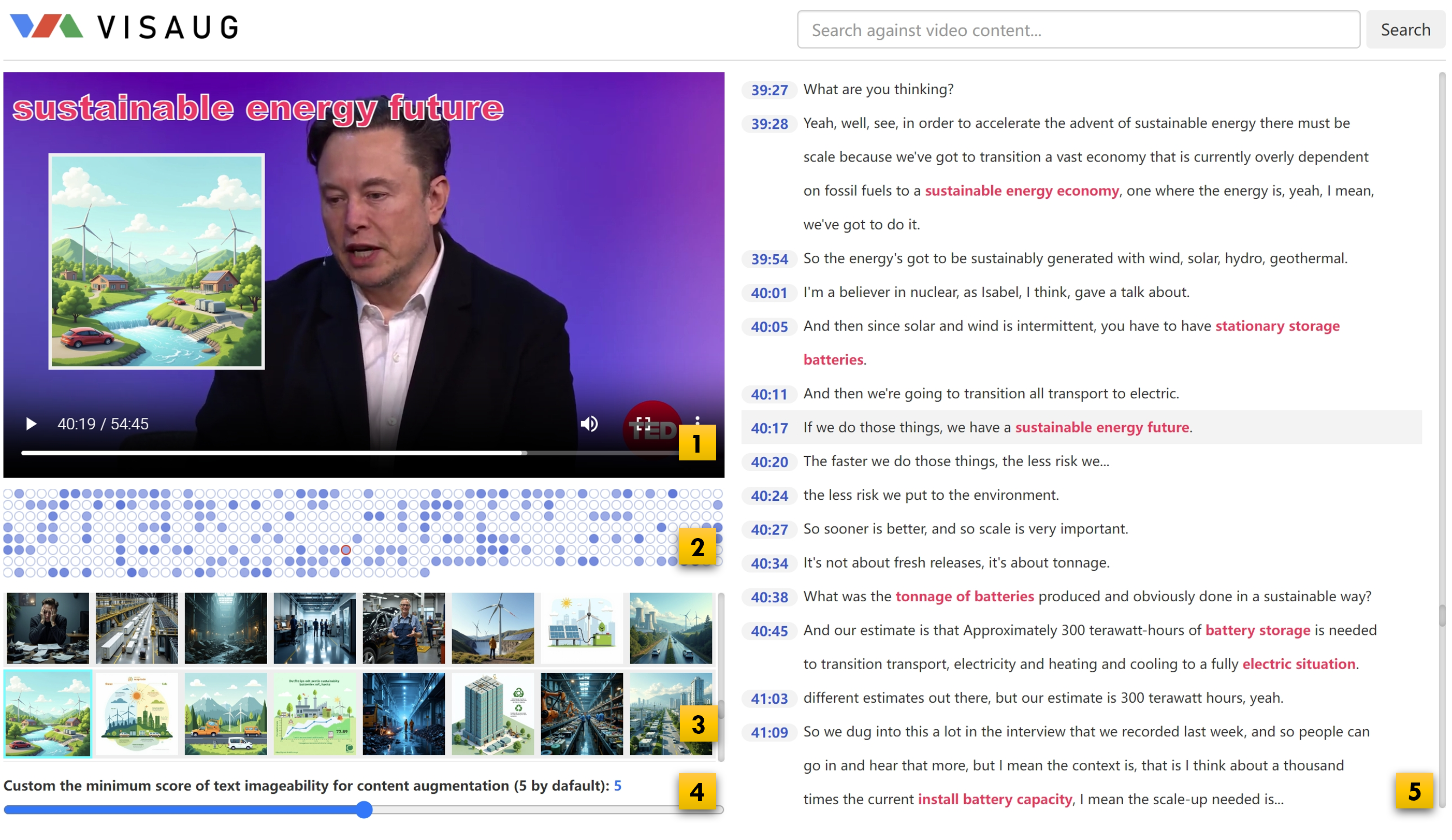}
    \caption{VisAug interface featuring five components: (1) video player with overlay visualizations; (2) transcript imageability visualization with color-coded segments; (3) visual augmentation storyboard gallery; (4) imageability threshold adjustment control; and (5) interactive transcript panel with highlighted key phrases for navigation.}
    \label{fig:ui}
\end{figure*}

\subsection{Augmentation Packing}

In the visual augmentation packing stage, the system leverages the saliency maps from the visual branch to determine optimal placement of the keyphrases and the generated image augmentations. This ensures that new visual elements are integrated seamlessly with the original video content, enhancing rather than disrupting the viewer's experience. For instance, augmentations might be positioned in less salient areas to provide additional information without obscuring key visual elements of the original video. The proposed method first manages image augmentation placement. It starts with the original size of the generated image and gradually reduces it while maintaining the original aspect ratio. For each size, it attempts to seek a location where less than a pre-defined number of pixels are salient within the image area. Upon finding a suitable spot, it returns the updated mask. The packing of text augmentations is performed similarly. These visual augmentations synchronize with video playing to facilitate content navigation, digestion, and engagement.

\section{User Interface of the System}
As illustrated in Figure~\ref{fig:ui}, the user interface of the proposed system mainly consists of five panels. The primary component is an embedded video player positioned at the top left of the interface (Panel \#1). This panel not only plays the video content but also serves as a canvas for dynamic overlay displays triggered by user interactions with the interface elements in Panel \#2 and \#3. Adjacent to the video player, on the right side, is the transcript panel (Panel \#5). This panel presents a time-stamped video transcript, with key phrases highlighted in red to draw attention to critical concepts and facilitate quick scanning of the video content.

Beneath the video player lies the transcript imageability visualization panel (Panel \#2).
The panel consists of a series of dots, each corresponding to a specific transcript segment, arranged in a distinctive zigzag pattern. This arrangement flows from top to bottom and left to right, following the chronological order of the video transcript. The zigzag layout allows for a compact representation of the entire video timeline within a limited vertical space. This design choice enables users to view a substantial portion of the video's structure at a glance, without the need for scrolling or expanding the panel. Each dot in this zigzag pattern is color-coded based on its associated transcript segment's imageability score, which ranges from 1 to 10. Dots representing segments with scores of 5 and above are rendered in color, while those below 5 are displayed in white. This color-coding scheme provides an immediate visual cue about the potential for rich visual content throughout the video's duration.
The zigzag arrangement, combined with the color-coding, creates a visually striking and information-rich timeline. Users can quickly scan the pattern to identify clusters of highly imageable content, potential segments of particular interest. This visualization approach enhances the user's ability to navigate and understand the video's content structure efficiently.
Moreover, the interactive nature of these dots allows users to hover over them to preview associated visual augmentations on the video player, or click to jump directly to the corresponding point in the video. This interactivity, coupled with the intuitive zigzag layout, makes the timeline visualization panel a powerful tool for non-linear content exploration and discovery.

The bottom of the interface houses the storyboard panel (Panel \#3), which displays a gallery of auto-generated image augmentations. These images serve as visual synopses of key moments or concepts within the video, offering users a quick visual summary of the content and enabling rapid navigation to points of interest. As users scroll through the images, they can hover over any augmentation to preview it on the video player or click to jump to the corresponding point in the video. User customization is facilitated through a range bar located at the bottom right of the interface (Panel \#4). This control allows users to adjust the minimum imageability score threshold for content augmentation, enabling them to tailor the density and frequency of visual aids to their preferences or needs. The visualization panel and the storyboard content will be dynamically refreshed to reflect the change of the minimum imageability score threshold.

The system incorporates sophisticated interaction design to enhance user engagement. When a user hovers their cursor over a dot in the timeline or an image in the storyboard, the corresponding visual augmentations, including keyphrases (rendered in red) and image augmentation (highlighted in white image border), are displayed as an overlay on the video player. This feature allows for seamless content preview without disrupting video playback. Furthermore, clicking on any transcript segment, storyboard image, or timeline dot prompts the video to begin playback from the associated timestamp, enabling precise navigation to specific points of interest within the video.

By integrating these components and interactions, VisAug creates a cohesive and intuitive interface that leverages AI-generated visual augmentations and large language models to significantly enhance the video watching experience. This system not only facilitates more efficient navigation of video content but also promotes deeper engagement and comprehension, particularly for complex or lengthy presentations. The multi-modal approach of VisAug, combining textual, visual, and interactive elements, caters to diverse user preferences, making it a powerful tool for both casual viewers and those engaged in more rigorous study or analysis of video content.

\section{Experiment}
To evaluate the effectiveness of the proposed VisAug system, we conducted a comprehensive user study examining its impact on navigation efficiency and user engagement with speech-rich video content. Through controlled task-based assessments, standardized questionnaires, and qualitative interviews, we systematically measured both objective performance metrics and subjective user experiences to provide robust validation of the system's capabilities.

\subsection{Experimental Setup}

\subsubsection{Task Design} Our evaluation used a rigorous experimental design within the subjects that compared VisAug with multiple baseline conditions, including standard video players, and the retrieval-based augmentation system Visual Caption~\cite{liu2023visual}. The positioning tasks involve locating video segments discussing AI social concerns, identifying content about government AI policy initiatives, and finding discussions about AI's impact on education. The participants in this study were asked to use the systems to complete the aforementioned positioning tasks. Following this, they completed a questionnaire assessing various aspects of their experience with the VisAug system and underwent a structured usability interview to gain a deeper understanding of their user experience and to gather improvement suggestions. All tasks are meticulously designed to ensure controlled data analysis while maintaining ecological validity, providing an environment that effectively tests system functionality under realistic conditions. This approach allows for robust evaluation of the VisAug system's ability to enhance video content navigation and user engagement.

\begin{figure}[!t]
\centering
\begin{tikzpicture}
   \begin{axis}[
       ybar,
       bar width=1.2cm,
       width=\columnwidth,
       height=5cm,
       ylabel={Task Accuracy (\%)},
       ymin=50,
       ymax=100,
       xtick={0,1,2},
       xticklabels={Control, {Visual Caption}, {VisAug (Ours)}},
       xticklabel style={align=center, font=\small},
       ylabel style={font=\small},
       major grid style={line width=.2pt,draw=gray!50},
       nodes near coords,
       nodes near coords align={vertical},
       enlarge x limits={abs=0.5},
       bar shift=0pt,
       axis lines*=left,
       tick align=outside,
   ]
   \addplot+[
       fill=gray!20,
       draw=gray!20,
       text=black
   ] coordinates {
       (0, 69.8)
   };
   \addplot+[
        fill=blue!20,
        draw=blue!20,
       text=black
   ] coordinates {
       (1, 72.4)
   };
   \addplot+[
       fill=purple!20,
        draw=purple!20,       
       text=black
   ] coordinates {
       (2, 82.3)
   };
   \end{axis}
\end{tikzpicture}
\caption{Task accuracy comparison across different methods.}
\label{fig:baseline_comparison}
\end{figure}
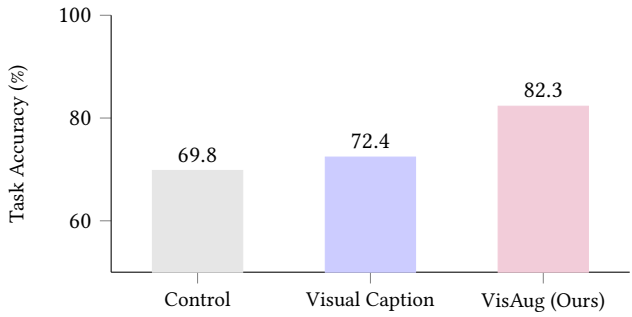

\subsubsection{Participant Recruitment}
We recruited 20 participants (12 female, 8 male) aged 20-30 with diverse academic backgrounds (2 doctoral, 7 master's, 11 undergraduate) from computer science, AI, interaction design, journalism, law, and product management. All participants demonstrated strong information processing capabilities and English proficiency. Each participant (labeled P1 to P20) completed the entire research process in approximately 1.5-2 hours, and received a \$15 reward as compensation at the end of the study.

\subsubsection{Experimental Procedure}
The study followed a structured four-phase protocol with condition randomization:
\begin{itemize}
    \item Training Phase: 10-minute system familiarization and environmental acclimation.
\item Task Completion: Participants completed  information location tasks per condition with performance metrics tracking.
\item Assessment: NASA-TLX workload assessment and System Usability Scale questionnaire.
\item Interview: 15-minute semi-structured interviews for qualitative feedback.
\end{itemize}

\subsection{Results and Analysis}

\subsubsection{Task Accuracy Performance}
Statistical analysis revealed significant performance advantages for VisAug across all baseline comparisons. Figure~\ref{fig:baseline_comparison} presents the results comparing VisAug against control and retrieval-based conditions. VisAug achieved 82.3\% task accuracy compared to 69.8\% for the control condition and 72.4\% for retrieval-based methods, representing improvements of 12.5\% and 9.9\%, respectively. The substantial improvement over the control condition demonstrates VisAug's effectiveness in enhancing user navigation capabilities beyond standard video interfaces. The significant advantage over existing retrieval-based methods validates our AIGC-based approach's superior contextual understanding and visual augmentation quality.

\begin{figure}[!t]
\centering
\begin{tikzpicture}
   \begin{axis}[
       ybar,
       bar width=0.8cm,
       width=\columnwidth,
       height=6cm,
       ylabel={Rating},
       ymin=4,
       ymax=7,
       xtick={0,1,2},
       xticklabels={{Semantic\\Matching}, {Visual\\Attractiveness}, {Information\\Gain}},
       xticklabel style={align=center, font=\small},
       ylabel style={font=\small},
       nodes near coords,
       nodes near coords align={vertical},
       enlarge x limits={abs=0.5},
       axis lines*=left,
       tick align=outside,
       legend style={at={(0.5,0.95)}, anchor=north, legend columns=2, font=\small},
   ]
   \addplot+[
       fill=blue!20,
       draw=blue!20,
       text=black
   ] coordinates {
       (0, 5.54)
       (1, 5.08)
       (2, 5.53)
   };
   \addlegendentry{Visual Caption}
   
   \addplot+[
       fill=purple!20,
       draw=purple!20,
       text=black
   ] coordinates {
       (0, 5.67)
       (1, 5.71)
       (2, 5.60)
   };
   \addlegendentry{VisAug (Ours)}
   \end{axis}
\end{tikzpicture}
\caption{Comparison of VisAug with Visual Caption across three evaluation metrics.}
\label{fig:retrieval_comparison}
\end{figure}
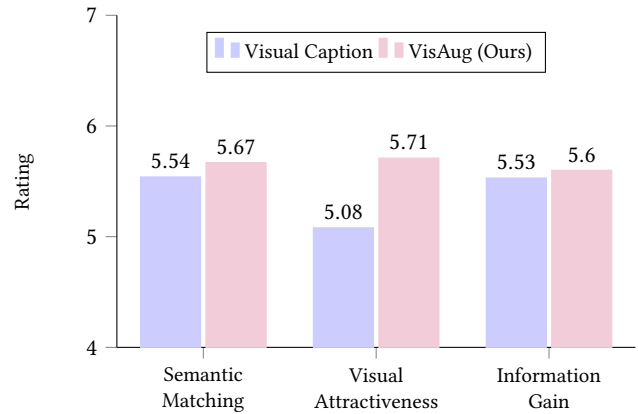

\subsubsection{User Preference}
To further validate these findings, we conducted a user preference study with 20 participants comparing AIGC-generated images with retrieval-based alternative across 100 video sentences. Participants were presented with paired comparisons and asked to select their preferred augmentation approach. We evaluated three key metrics:
\begin{itemize}
\item \textbf{Semantic Matching} measures the alignment between visual content and the semantic meaning of corresponding speech segments, evaluating how accurately the augmentation represents discussed concepts and themes.
\item \textbf{Visual Attractiveness} assesses the aesthetic appeal and visual quality of augmentation content, including factors such as image clarity, composition, and overall visual appeal that contribute to user engagement.
\item \textbf{Information Gain} evaluates whether visual augmentation provides additional value beyond speech content alone, measuring how effectively visual elements help users comprehend complex concepts or gain supplementary understanding.
\end{itemize}
These metrics collectively provide a holistic assessment of augmentation quality, ensuring that our system not only generates contextually relevant content but also delivers visually appealing and informationally valuable enhancements to users. Our primary comparison against Visual Caption, a state-of-the-art retrieval-based system, demonstrates consistent improvements across key performance metrics, as shown in Figure~\ref{fig:retrieval_comparison}. The 81\% preference for AIGC-generated content validates our approach's superior contextual relevance and semantic alignment compared to retrieval-based methods. This substantial preference margin indicates that users consistently recognize and value the enhanced contextual appropriateness and visual quality of generated content over retrieved alternative. The improved semantic matching scores (5.67 vs 5.54) and significantly higher visual attractiveness ratings (5.71 vs 5.08) demonstrate that AIGC-based augmentation provides more contextually relevant and aesthetically appealing visual enhancements for speech-rich video content.

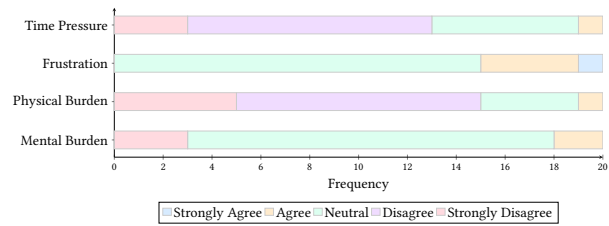
\begin{figure}[!t]
\centering
\begin{tikzpicture}[scale=0.45]
\definecolor{softrose}{RGB}{255,218,224}
\definecolor{lavenderblush}{RGB}{240,220,255}
\definecolor{mintcream}{RGB}{220,255,240}
\definecolor{peachpuff}{RGB}{255,235,205}
\definecolor{powderblue}{RGB}{215,235,255}
\begin{axis}[
   width=0.9\textwidth,
   height=6cm,
   xbar stacked,
   bar width=15pt,
   xmin=0,
   xmax=20,
   xlabel={Frequency},
   xlabel style={font=\LARGE},
   ytick={1,2,3,4},
   yticklabels={Mental Burden, Physical Burden, Frustration, Time Pressure},
   yticklabel style={font=\LARGE},
   legend style={at={(0.5,-0.3)}, anchor=north, legend columns=5, font=\LARGE},
   reverse legend=true,
   axis x line=bottom,
   axis y line=left,
   enlarge y limits=0.15,
]
\addplot[fill=softrose, draw=gray!40] coordinates {
   (3,1)
   (5,2)
   (0,3)
   (3,4)
};
\addplot[fill=lavenderblush, draw=gray!40] coordinates {
   (0,1)
   (10,2)
   (0,3)
   (10,4)
};
\addplot[fill=mintcream, draw=gray!40] coordinates {
   (15,1)
   (4,2)
   (15,3)
   (6,4)
};
\addplot[fill=peachpuff, draw=gray!40] coordinates {
   (2,1)
   (1,2)
   (4,3)
   (1,4)
};
\addplot[fill=powderblue, draw=gray!40] coordinates {
   (0,1)
   (0,2)
   (1,3)
   (0,4)
};
\legend{Strongly Disagree, Disagree, Neutral, Agree, Strongly Agree}
\end{axis}
\end{tikzpicture}
\caption{Task Load Index showing user perception of mental and physical demands.}
\label{fig:taskload}
\end{figure}

\begin{figure}[!t]
\centering
\begin{tikzpicture}[scale=0.45]
\definecolor{softrose}{RGB}{255,218,224}
\definecolor{lavenderblush}{RGB}{240,220,255}
\definecolor{mintcream}{RGB}{220,255,240}
\definecolor{peachpuff}{RGB}{255,235,205}
\definecolor{powderblue}{RGB}{215,235,255}
\begin{axis}[
   width=\textwidth,
   height=4cm,
   xbar stacked,
   bar width=15pt,
   xmin=0,
   xmax=20,
   xlabel={Frequency},
   xlabel style={font=\LARGE},
   ytick={1,2},
   yticklabels={Effort, Efficacy},
   yticklabel style={font=\LARGE},
   legend style={at={(0.5,-0.5)}, anchor=north, legend columns=5, font=\LARGE},
   reverse legend=true,
   axis x line=bottom,
   axis y line=left,
   enlarge y limits=0.3,
]
\addplot[fill=softrose, draw=gray!40] coordinates {
   (2,1)
   (2,2)
};
\addplot[fill=lavenderblush, draw=gray!40] coordinates {
   (4,1)
   (0,2)
};
\addplot[fill=mintcream, draw=gray!40] coordinates {
   (10,1)
   (6,2)
};
\addplot[fill=peachpuff, draw=gray!40] coordinates {
   (3,1)
   (10,2)
};
\addplot[fill=powderblue, draw=gray!40] coordinates {
   (1,1)
   (2,2)
};
\legend{Strongly Disagree, Disagree, Neutral, Agree, Strongly Agree}
\end{axis}
\end{tikzpicture}
\caption{The level of engagement \& utility of the system as perceived by users.}
\label{fig:engagement}
\end{figure}
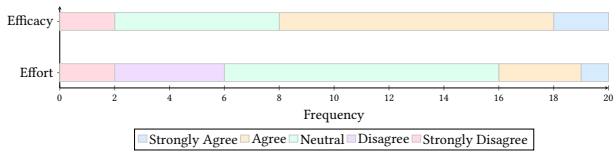

\begin{figure}[!t]
\centering
\begin{tikzpicture}[scale=0.45]
\definecolor{softrose}{RGB}{255,218,224}
\definecolor{lavenderblush}{RGB}{240,220,255}
\definecolor{mintcream}{RGB}{220,255,240}
\definecolor{peachpuff}{RGB}{255,235,205}
\definecolor{powderblue}{RGB}{215,235,255}
\begin{axis}[
   width=0.9\textwidth,
   height=6cm,
   xbar stacked,
   bar width=15pt,
   xmin=0,
   xmax=20,
   xlabel={Frequency},
   xlabel style={font=\LARGE},
   ytick={1,2,3,4},
   yticklabels={Understanding, Clarification, Engagement, New Thinking},
   yticklabel style={font=\LARGE},
   legend style={at={(0.5,-0.3)}, anchor=north, legend columns=5, font=\LARGE},
   reverse legend=true,
   axis x line=bottom,
   axis y line=left,
   enlarge y limits=0.15,
]
\addplot[fill=softrose, draw=gray!40] coordinates {
   (1,1)
   (2,2)
   (1,3)
   (0,4)
};
\addplot[fill=lavenderblush, draw=gray!40] coordinates {
   (5,1)
   (4,2)
   (2,3)
   (9,4)
};
\addplot[fill=mintcream, draw=gray!40] coordinates {
   (0,1)
   (0,2)
   (7,3)
   (3,4)
};
\addplot[fill=peachpuff, draw=gray!40] coordinates {
   (8,1)
   (14,2)
   (4,3)
   (6,4)
};
\addplot[fill=powderblue, draw=gray!40] coordinates {
   (6,1)
   (0,2)
   (6,3)
   (2,4)
};
\legend{Strongly Disagree, Disagree, Neutral, Agree, Strongly Agree}
\end{axis}
\end{tikzpicture}
\caption{Effect of Visual Enhancement on learning and cognitive processes.}
\label{fig:enhancement}
\end{figure}
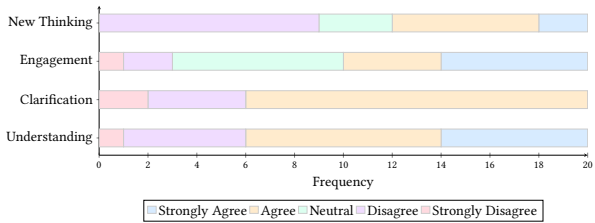

\subsubsection{User Experience Feedback}
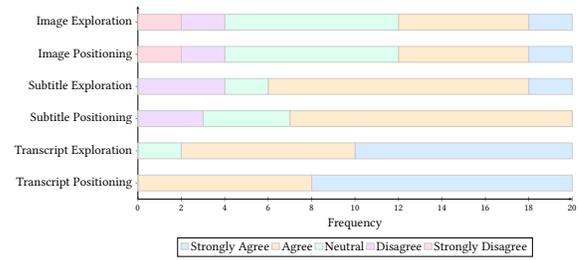
\begin{figure}[!t]
\centering
\begin{tikzpicture}[scale=0.4]
\definecolor{softrose}{RGB}{255,218,224}
\definecolor{lavenderblush}{RGB}{240,220,255}
\definecolor{mintcream}{RGB}{220,255,240}
\definecolor{peachpuff}{RGB}{255,235,205}
\definecolor{powderblue}{RGB}{215,235,255}
\begin{axis}[
   width=0.9\textwidth,
   height=8cm,
   xbar stacked,
   bar width=15pt,
   xmin=0,
   xmax=20,
   xlabel={Frequency},
   xlabel style={font=\LARGE},
   ytick={1,2,3,4,5,6},
   yticklabels={Transcript Positioning, Transcript Exploration, Subtitle Positioning, Subtitle Exploration, Image Positioning, Image Exploration},
   yticklabel style={font=\LARGE},
   legend style={at={(0.5,-0.2)}, anchor=north, legend columns=5, font=\LARGE},
   reverse legend=true,
   axis x line=bottom,
   axis y line=left,
   enlarge y limits=0.1,
]
\addplot[fill=softrose, draw=gray!40] coordinates {
   (0,1)
   (0,2)
   (0,3)
   (0,4)
   (2,5)
   (2,6)
};
\addplot[fill=lavenderblush, draw=gray!40] coordinates {
   (0,1)
   (0,2)
   (3,3)
   (4,4)
   (2,5)
   (2,6)
};
\addplot[fill=mintcream, draw=gray!40] coordinates {
   (0,1)
   (2,2)
   (4,3)
   (2,4)
   (8,5)
   (8,6)
};
\addplot[fill=peachpuff, draw=gray!40] coordinates {
   (8,1)
   (8,2)
   (13,3)
   (12,4)
   (6,5)
   (6,6)
};
\addplot[fill=powderblue, draw=gray!40] coordinates {
   (12,1)
   (10,2)
   (0,3)
   (2,4)
   (2,5)
   (2,6)
};
\legend{Strongly Disagree, Disagree, Neutral, Agree, Strongly Agree}
\end{axis}
\end{tikzpicture}
\caption{Usage of Features showing participant preferences across system functionalities.}
\label{fig:features}
\end{figure}

\begin{figure}[!t]
\centering
\begin{tikzpicture}[scale=0.45]
\definecolor{softrose}{RGB}{255,218,224}
\definecolor{lavenderblush}{RGB}{240,220,255}
\definecolor{mintcream}{RGB}{220,255,240}
\definecolor{peachpuff}{RGB}{255,235,205}
\definecolor{powderblue}{RGB}{215,235,255}
\begin{axis}[
   width=0.9\textwidth,
   height=5cm,
   xbar stacked,
   bar width=15pt,
   xmin=0,
   xmax=20,
   xlabel={Frequency},
   xlabel style={font=\LARGE},
   ytick={1,2,3},
   yticklabels={Task Difficulty, Time Spent, Support Information},
   yticklabel style={font=\LARGE},
   legend style={at={(0.5,-0.4)}, anchor=north, legend columns=5, font=\LARGE},
   reverse legend=true,
   axis x line=bottom,
   axis y line=left,
   enlarge y limits=0.2,
]
\addplot[fill=softrose, draw=gray!40] coordinates {
   (0,1)
   (0,2)
   (1,3)
};
\addplot[fill=lavenderblush, draw=gray!40] coordinates {
   (0,1)
   (0,2)
   (0,3)
};
\addplot[fill=mintcream, draw=gray!40] coordinates {
   (3,1)
   (4,2)
   (2,3)
};
\addplot[fill=peachpuff, draw=gray!40] coordinates {
   (5,1)
   (6,2)
   (12,3)
};
\addplot[fill=powderblue, draw=gray!40] coordinates {
   (12,1)
   (10,2)
   (5,3)
};
\legend{Strongly Disagree, Disagree, Neutral, Agree, Strongly Agree}
\end{axis}
\end{tikzpicture}
\caption{After-Scenario Questionnaire results measuring overall system satisfaction.}
\label{fig:after}
\end{figure}
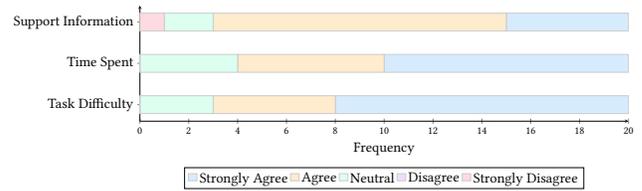

Participants completed a comprehensive questionnaire assessing various aspects of their experience with the VisAug system. Figure~\ref{fig:taskload} shows participants' assessment of task load dimensions, indicating moderate mental demand but relatively low frustration levels, suggesting the system was engaging without being overwhelming.
Engagement metrics revealed that most participants found the system effective, as shown in Figure~\ref{fig:engagement}. A majority of respondents rated the system as efficient and effective, with minimal additional effort required for operation.
The effectiveness of visual enhancements was particularly notable in clarifying complex concepts and increasing engagement, as illustrated in Figure~\ref{fig:enhancement}. This suggests that the system's visual elements successfully augmented the verbal content.
Feature usage analysis (Figure~\ref{fig:features}) revealed interesting patterns, with transcript-based features receiving the strongest positive ratings, followed by subtitle axis features. Image storyboard features showed more mixed responses, indicating potential areas for improvement.
Overall satisfaction metrics (Figure~\ref{fig:after}) demonstrated positive outcomes, with most participants reporting satisfaction with supporting information and appropriate time requirements.
\subsubsection{Qualitative Interview Analysis}
Interview data revealed four key system components:

\paragraph{Subtitle Panel.} Users valued the panel for comprehension support, particularly with non-native content (P2, P3, P10). Usability issues included automatic scrolling constraints (P11, P12) and requests for customizable controls (P15). Suggestions included mouseover image enhancements (P17) and improved marker visibility (P19). P19 highlighted subtitles' value for quick content overview.

\paragraph{Keyword Highlighting.} While generally helpful for navigation, users noted term selection relevance issues (P7, P11). Common terms were highlighted too frequently, while contextually important phrases were missed (P20). Suggestions included user-customizable keywords (P5), improved algorithmic selection, and context summaries (P14). P12 valued the feature for identifying key content and emotional tone.

\paragraph{Navigation.} Users requested topic-based segmentation (P2, P3, P9) and annotation capabilities (P7). P11 suggested a centered video layout with peripheral elements to address attention dispersion. Most relied heavily on subtitles and keywords, with P12 requesting enhanced timestamp functionality.

\paragraph{Image Enhancement.} Mixed feedback revealed limited utility for abstract concepts (P2, P3, P5). Users suggested event-based visualization (P9), data-specific charts (P10, P15), and context-appropriate imagery (P11, P12). P13 noted tone mismatches where cartoonish images undermined serious topics. In general, users sought more contextually relevant visuals aligned with the tone of the content.

\section{Conclusion}

In this paper, we have presented VisAug, a novel system for enhancing speech-rich web videos through intelligent visual augmentations. Our approach introduces a dual-channel processing pipeline that synergistically combines advanced computer vision techniques with sophisticated natural language processing methods to create a more engaging and informative video viewing experience.
A significant contribution of our work lies in the utilization of AI-generated visual augmentations, which offer substantial advantages over existing retrieval-based methods. Unlike traditional approaches that rely on pre-existing image databases, our context-aware prompt formulation scheme, coupled with state-of-the-art text-to-image generation models, allows for the creation of bespoke visual content that precisely aligns with the video's narrative.  This AI-driven approach excels in capturing the multifaceted nature of imageable content in video transcripts, offering several key benefits, including contextual accuracy, conceptual flexibility, and dynamic adaptation. These advantages enable VisAug to provide a more comprehensive, engaging, and informative viewing experience that captures the full complexity and nuance of speech-rich video transcript.

\begin{acks}
 This work was supported by Guangdong Basic and Applied Basic Research Foundation (2023A1515011639).
\end{acks}

\bibliographystyle{ACM-Reference-Format}
\balance
\bibliography{sample-base}


\end{document}